\documentclass[twocolumn,preprintnumbers,amsmath,superscriptaddress,amssymb,aps]{revtex4-1}

\usepackage{graphicx}
\usepackage{dcolumn}
\usepackage{bm}
\usepackage{amsmath}
\usepackage{amssymb}
\usepackage{graphicx}
\usepackage{indentfirst}
\usepackage{booktabs}
\usepackage{multirow}
\usepackage{colortbl}

\selectfont

\begin{document}

\title{Large Second-Harmonic Generation and Linear Electro-Optic Effect in Trigonal Selenium and Tellurium}

\author{Meijuan Cheng}
\address{Department of Physics, Collaborative Innovation Center for Optoelectronic Semiconductors
and Efficient Devices, Key Laboratory of Low Dimensional Condensed Matter Physics
(Department of Education of Fujian Province), Jiujiang Research Institute, Xiamen University, Xiamen 361005, China}
\author{Shunqing Wu}
\address{Department of Physics, Collaborative Innovation Center for Optoelectronic Semiconductors
and Efficient Devices, Key Laboratory of Low Dimensional Condensed Matter Physics
(Department of Education of Fujian Province), Jiujiang Research Institute, Xiamen University, Xiamen 361005, China}
\author{Zi-Zhong Zhu}
\email{zzhu@xmu.edu.cn}
\address{Department of Physics, Collaborative Innovation Center for Optoelectronic Semiconductors
and Efficient Devices, Key Laboratory of Low Dimensional Condensed Matter Physics
(Department of Education of Fujian Province), Jiujiang Research Institute, Xiamen University, Xiamen 361005, China}
\address{Fujian Provincial Key Laboratory of Theoretical and Computational Chemistry, Xiamen 361005, China}
\author{Guang-Yu Guo}
\email{gyguo@phys.ntu.edu.tw}
\address{Department of Physics and Center for Theoretical Physics, National Taiwan University, Taipei 10617, Taiwan}
\address{Physics Division, National Center for Theoretical Sciences, Hsinchu 30013, Taiwan}


\date{\today}

\begin{abstract}
Trigonal selenium and tellurium crystalize in helical chain-like structures and thus
possess interesting properties such as current-induced spin polarization, gyrotropic effects and  nonlinear optical responses.
By performing systematic {\it ab initio} calculations based on density functional theory in the generalized gradient approximation
plus scissors correction, we study their linear and nonlinear optical (NLO) properties.
We find that the two materials exhibit significant second-harmonic generation (SHG) and linear electro-optic (LEO) effect.
In particular, the SHG coefficients ($\chi^{(2)}$) of tellurium are huge in the photon energy range of 0$\sim$3 eV,
with the magnitudes of $\chi^{(2)}_{xxx}$  being as large as 3640 pm/V, which are about sixteen times larger than
that of GaN, a widely used NLO material. 
On the other hand, trigonal selenium is found to possesses large LEO coefficients $r_{xxx}(0)$ which are
more than six times larger than that of GaN.
Therefore, trigonal tellurium and selenium may find valuable applications in NLO and LEO devices such
as frequency conversion, electro-optical switch and light signal modulator.
The energy positions and shapes of the principal features in the calculated optical dielectric functions
of both materials agree rather well with the available experimental ones, and are also analyzed in terms
of the calculated band structures especially symmetries of the involved band states and
dipole transition selection rules.
\end{abstract}

\maketitle


\section{INTRODUCTION}
The interaction between the intense optical fields and materials may induce strong nonlinear optical (NLO) responses.\cite{Shen1984,Boyd2003}
Noncentrosymmetric materials with large second-order nonlinear optical susceptibility ($\chi^{(2)}$) play a crucial role in the development
of modern optical and electro-optical devices such as lasers, frequency conversions, electro-optic modulator and switches.\cite{Boyd2003}
Second harmonic generation (SHG), a special case of sum frequency generation, is perhaps the best-known NLO effect.
Since 1960s, the SHG has been investigated extensively in bulk 
semiconductors\cite{Boyd2003,Chang1965,zhong1993,Hughes1996,Gavrilenko2000,Cai2009} and more recently also in one dimensional
(see, e.g., \cite{guo2004,guo2005} and references therein)
and two-dimensional (see, e.g., \cite{wang2015,hu2017,wang2017,panday2018} and references therein) materials.
Furthermore, because of its high sensitivity to local structure symmetry and also its absence in materials with inversion symmetry,
the SHG has been a powerful probe of surfaces and interfaces.\cite{Shen1984}
Linear electro-optic (LEO) effect, another second-order electric polarization response of a NLO material, 
refers to the linear refractive index variation ($\Delta n$) with the applied electric field strength ($E$),
$\Delta n = n^3rE/2$, where $n$ is the refraction index and $r$ the LEO coefficient.\cite{Boyd2003}
The LEO effect thus allows one to use an electrical signal to control the amplitude, phase or direction
of light beam in the NLO material, and leads to a widely used means for high-speed optical modulation and
sensing devices. 

Since their discovery, trigonal selenium and tellurium have attracted considerable attention
due to their unique properties, such as a high degree of anisotropy, broken inversion symmetry and chirality.
Trigonal selenium has valuable technological applications such as rectifiers, photocells, photographic exposure meters
and xerography to medical diagnostics due to their photoconductivity
in the entire visible range\cite{smith1980,mees1995,johnson1999,fan2005}.
As a narrow band gap semiconductor, tellurium possesses excellent thermoelectric property \cite{peng2014,lin2016}
and thus can be used for thermoelectrics. Furthermore, tellurium also exhibits such interesting behaviors as
current-induced spin polarization\cite{shalygin2012},
trivial insulator to strong topological insulator transition under shear strain\cite{agapito2013},
circular photon drag effect\cite{shalygin2016},
robust control over current-induce magnetic torques\cite{csahin2018},
and gyrotropic effects\cite{tsirkin2018}.
Finally, their electronic band structures host Dirac and Weyl nodes due to the broken inversion symmetry
and spin-orbit coupling (SOC)\cite{Hirayama2015,nakayama2017}.

As semiconductors without inversion symmetry, selenium and tellurium exhibit nonlinear optical responses.
Although their optical properties\cite{patel1966,tutihasi1967,roberts1968,mcfee1970,day1971,zhong1993}
have been extensively studied, their nonlinear optical effects have hardly been investigated.
Only the second-order NLO susceptibility
element $\chi^{(2)}_{xxx}$ has been measured at the CO$_2$ laser frequency ($\hbar\omega = 0.113$ meV)
for selenium and tellurium\cite{patel1966,mcfee1970,day1971}
and calculated at zero frequency for selenium\cite{zhong1993}.
No study on the other nonzero element $\chi^{(2)}_{xyz}$ has been reported.
Since the SHG depends on the electronic band structure, dipole transition matrix and specific frequency
and orientation of the applied optical field, it is of interest to know all the nonzero SHG elements
over the entire optical frequency range. The LEO effect in these two materials has been investigated either.
Therefore, the main objectives of this paper are as follows. Firstly, we want to perform systematic {\it ab initio} calculations
of all the nonzero elements of the second-order NLO susceptibility tensor $\chi^{(2)}_{\alpha\beta\gamma}(-2\omega,\omega,\omega)$
of both materials for the whole optical frequency range. The results will tell us whether trigonal selenium and tellurium 
are promising NLO materials for optical and opto-electronic devices. Secondly, we also want to calculate  
linear optical dielectric functions $\varepsilon(\omega)$ for both selenium and tellurium in order to
understand the interesting observed optical phenomena. The calculated dielectric functions 
will also help us to understand the obtained SHG coefficients.~\cite{Gavrilenko2000,guo2005,wang2015} 
Furthermore, since a NLO material with a large LEO coefficient needs to simultaneously posses a large
second-order NLO susceptibility and a low dielectric constant, the dielectric functions are also
required for evaluating the LEO coefficient. 
Our findings are expected to stimulate further experimental
investigations on the NLO properties of these two interesting helical chain-like materials.

The paper is organized as follows. In sec. \uppercase\expandafter{\romannumeral2},
we present the theoretical approach and computational details. In sec. \uppercase\expandafter{\romannumeral3},
the calculated optical dielectric function, SHG and LEO coefficients over the entire optical frequency range are presented.
Comparison of the obtained SHG and LEO coefficients of the two materials with the knowned NLO materials
suggests that they are superior NLO materials. The theoretical SHG coefficients 
are compared with the measured values at the CO$_2$ laser frequency, and also analyzed in terms
of one- and two-photon resonances via the calculated absorptive parts of the dielectric function.
Furthermore, we compare the calculated dielectric function over the whole optical frequency range 
with the available experimental results and also analyze their interesting features
in terms of the symmetry of band states at high symmetry $k$-points in the
Brillouin zone. Finally, a summery of this work is given in sec. \uppercase\expandafter{\romannumeral4}.

\section{STRUCTURE AND COMPUTATIONAL METHOD }

The crystal structure of trigonal selenium and tellurium\cite{teuchert1975,Keller1997} is schematically shown in Fig. 1.
It consists of the helical chains arranged in a hexagonal array\cite{teuchert1975,Keller1997}. 
The three atoms in the unit cell are situated at positions ($u, 0, 0$), ($0, u, 1/3$) and ($-u, -u, 2/3$).
The space group is either $P3_121$ ($D_3^4$) or $P3_221$ ($D_3^6$) depending on whether
it has the right-handed or left-handed screw. Nonetheless, the two different helical structures are related to each
other by spatial inversion. Thus their energy bands and linear optical properties should be identical.
Furthermore, their nonzero elements of the second-order NLO susceptibility tensor would be the same
and are also related to each other, as will be explained in sec. III(c) below. 
Therefore, we only consider the $D_3^4$ case here.
The structures could be viewed as being derived from the Peierls distortion\cite{seifert1995}
of the six-coordinated simple cubic structure. The valence electron configurations of selenium and tellurium
are $4s^2$$4p^4$ and $5s^2$$5p^4$, respectively, i.e., the one third $p$ bands are empty. Therefore,
every atom is covalently bonded to two neighboring atoms along each chain, and interacts with four second
near-neighbor atoms of the adjacent chains by van der Waals forces. The intrachain bonding and interchain bonding
correspond to the $p$ bonding and the lone pair states, respectively. The overlap of electronic orbitals
arising from the lone pair and antibonding states on neighbouring chains induces the covalent bond stretched
to infinity. Furthermore, such overlap and the repellant exchange interaction between lone pair orbitals
stabilize the helical chains. As shown in Fig. 1, lattice constant $a$ amounts to the
distance between the adjacent chains. Lattice constant $c$ is equal to the height of the unit cell
in the chain direction (the $c$ axis). Interestingly, the atomic position parameter $u = q/a$ can be related to
lattice constant $a$ and the helix radius $q$.
Note that the ratio ($R/r$) of the interchain and intrachain distances of tellurium is smaller than
that of selenium (Table I), thus implying that tellurium would exhibit a weaker structural anisotropy.

\begin{figure}[htb]
\begin{center}
\includegraphics[width=6.0cm]{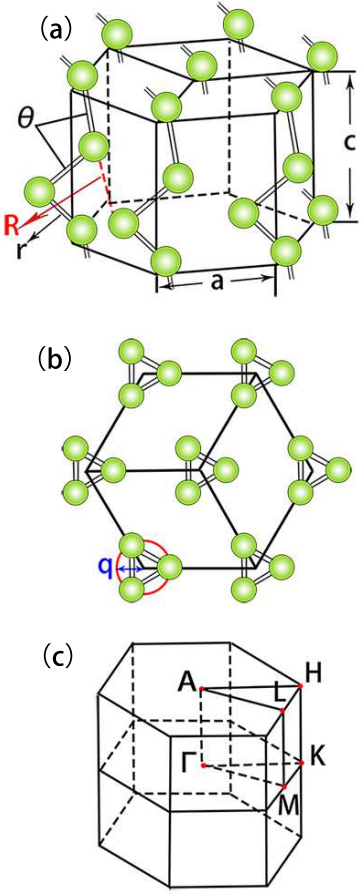}
\end{center}
\caption{(a) Side and (b) top views of the trigonal crystalline structure of selenium and tellurium
as well as (c) the associated Brillouin zone. $a$ and $c$ are lattice constants.
$\theta$, $r$ and $R$ are bond angle, intrachain and interchain distances, respectively.
$q$ denotes the radius of the helices.}
\end{figure}

\begin{table}
\caption{Calculated (the) and measured (exp) lattice constants $a$ and $c$, atomic-position parameter $u$,
the distances of intrachain $r$ and interchain $R$, bond angle $\theta$, cell volume $V$.
}
\begin{tabular}{c c c c c c c c c}
        &    & $a$ (\AA) & $c$ (\AA) & $u$ & $r$ (\AA) & $R$ (\AA) & $\theta (^{\circ})$ &   $V$ (\AA$^3$) \\  \hline
  Se  & exp\footnotemark[1] & 4.3662 & 4.9536 & 0.2254 & 2.3732 & 3.4358 & 103.07 & 81.78 \\
      & the                 & 4.3084 & 5.0874 & 0.2315 & 2.4207 & 3.3958 & 103.66 & 81.78 \\
  Te  & exp\footnotemark[2] & 4.4511 & 5.9262 & 0.2633 & 2.8307 & 3.4919 & 103.27 & 101.67 \\
      & the                 & 4.4345 & 5.9707 & 0.2754 & 2.9044 & 3.4428 & 101.79 & 101.67 \\
\hline
\end{tabular}

\footnotemark[1]{Reference~\onlinecite{teuchert1975}.}

\footnotemark[2]{Reference~\onlinecite{Keller1997}.}
\end{table}

The present {\it ab initio} calculations are performed based on the density functional theory. The exchange correlation
approximation is treated with the generalized gradient
approximation (GGA) of Perdew, Burke, and Ernzerhof\cite{perdew1996}.
The highly accurate full-potential projector-augmented wave (PAW) method\cite{blochl1994},
as implemented in the VASP package\cite{kresse1996ab,kresse1996},
is used. A large plane-wave cutoff of 450 eV is adopted throughout. The PAW potentials are used to describe
the electron-ion interaction, with 6 valence electrons for Se ($4s^2 4p^4$) and also 6 for Te ($5s^2 5p^4$).
The structural optimization and self-consistent electronic structure calculations
are performed by using the tetrahedron method~\cite{blochl1994tetrahedron} with a $k$-point mesh of 10$\times$10$\times$8.
The density of states (DOS) is evaluated from the self-consistent band structure with 
a much denser $k$-point mesh of 20$\times$20$\times$18 for Se and of 20$\times$20$\times$16
for Te.

We first perform structural optimizations of the atomic positions and lattice constants with the conjugate gradient technique.
As shown in Table I, the calculated lattice constants agree rather well with that of experiments\cite{teuchert1975,Keller1997}
except lattice constant $c$ of selenium which is about 2.6 \% too small. Nonetheles, the calculated
linear and nonlinear optical spectra using the theoretical structural parameters look almost the same as that obtained
using the experimental structural parameters.
Therefore, only the results calculated by using the experimental structural parameters are presented in this paper.

In this work, the linear optical dielectric function and nonlinear optical susceptibility are calculated based
on the linear response formalism with the independent-particle approximation, as described previously\cite{guo2004,guo2005ab}.
Therefore, the imaginary (absorptive) part of the dielectric function $\varepsilon(\omega)$ due to direct
interband transition is given by the Fermi golden rules\cite{guo2004,guo2005ab},
\begin{equation}
\varepsilon_{a}'' (\omega) = \frac{4\pi^2}{\Omega\omega^2}
\sum_{i\in VB,j\in CB}\sum_{k}w_{k}|p_{ij}^{a}|^{2}
\delta(\epsilon_{{\bf k}j}-\epsilon_{{\bf k}i}-\omega),
\end{equation}
where $\omega$  is  the photon energy and $\Omega$  is the unit cell volume. VB and CB represent the valence and conduction bands, respectively.
The dipole transition matrix elements $p_{ij}^{a} = \langle\textbf{k}\emph{j}|\hat{p}_{a}|\textbf{k}i\rangle$
are obtained from the self-consistent band structures within the PAW formalism.\cite{Adolph2001}
Here $|${\bf k}$n\rangle$ is the $n$th Bloch state wave function with crystal momentum {\bf k}, and $a$ denotes the cartesian component.
The real (dispersive) part of the dielectric function is then obtained from the calculated $\varepsilon''(\omega)$
by the Kramer-Kronig transformation\cite{guo2004,guo2005ab},
\begin{equation}
\varepsilon'(\omega) = 1+\frac{2}{\pi}{\bf P} \int _{0}^{\infty }
d\omega ^{'}\frac{\omega '\varepsilon'' (\omega ')}{\omega^{'2}-\omega ^{2}}.
\end{equation}
Here ${\bf P}$ represents the principal value of the integral.

The imaginary part of the second-order optical susceptibility due to direction interband transitions is given by \cite{guo2004,guo2005}
\begin{equation}
\chi''^{(2)}_{abc}(-2\omega,\omega,\omega) = \chi''^{(2)}_{abc,VE}(-2\omega,\omega,\omega)+\chi''^{(2)}_{abc,VH}(-2\omega,\omega,\omega),
\end{equation}
where the contribution due to the so-called virtual- electron (VE) process is\cite{guo2004,guo2005}
\begin{equation}
\begin{split}
\chi''^{(2)}_{abc,VE} = -\frac{\pi}{2\Omega}\sum_{i\in VB}\sum_{j,l\in CB}\sum_{\bf k}w_{\bf k}
\{\frac{Im[p_{jl}^{a}\langle p_{li}^{b} p_{ij}^{c}\rangle]}{\epsilon_{li}^{3}(\epsilon_{li}+\epsilon_{ji})}\delta(\epsilon_{li}-\omega) \\
-\frac{Im[p_{ij}^{a}\langle p_{jl}^{b} p_{li}^{c}\rangle]}{\epsilon_{li}^{3}(2\epsilon_{li}-\epsilon_{ji})}\delta(\epsilon_{li}-\omega)
+\frac{16Im[p_{ij}^{a}\langle p_{jl}^{b} p_{li}^{c}\rangle]}{\epsilon_{ji}^{3}(2\epsilon_{li}^{3}-\epsilon_{ji}^{3})}\delta(\epsilon_{ji}-2\omega)\},
\end{split}
\end{equation}
and that due to the virtual-hole (VH) process is\cite{guo2004,guo2005}
\begin{equation}
\begin{split}
\chi''^{(2)}_{abc,VH} = \frac{\pi}{2\Omega}\sum_{i,l\in VB}\sum_{j\in CB}\sum_{\bf k}w_{\bf k}
\{\frac{Im[p_{li}^{a}\langle p_{ij}^{b} p_{jl}^{c}\rangle]}{\epsilon_{jl}^{3}(\epsilon_{jl}+\epsilon_{ji})}\delta(\epsilon_{jl}-\omega) \\
-\frac{Im[p_{ij}^{a}\langle p_{jl}^{b} p_{li}^{c}\rangle]}{\epsilon_{jl}^{3}(2\epsilon_{jl}-\epsilon_{ji})}\delta(\epsilon_{jl}-\omega)
+\frac{16Im[p_{ij}^{a}\langle p_{jl}^{b} p_{li}^{c}\rangle]}{\epsilon_{ji}^{3}(2\epsilon_{jl}-\epsilon_{ji})}\delta(\epsilon_{ji}-2\omega)\}.
\end{split}
\end{equation}
Here $\epsilon_{ji}$ = $\epsilon_{kj}$-$\epsilon_{ki}$ and $\langle p_{jl}^{b}p_{li}^{c}\rangle = \frac{1}{2}(p_{jl}^{b}p_{li}^{c}+p_{li}^{b}p_{jl}^{c})$.
The real part of the second-order optical susceptibility is then obtained from the calculated $\chi''^{(2)}_{abc}$
by the Kramer-Kronig transformation\cite{guo2004,guo2005}
\begin{equation}
\chi'^{(2)}(-2\omega,\omega,\omega) = \frac{2}{\pi}{\bf P} \int_0^{\infty}d\omega'
\frac{\omega'\chi''^{(2)}(2\omega',\omega',\omega')}{\omega'^2-\omega^2}.
\end{equation}
The linear electro-optic coefficient $r_{abc}(\omega)$ is related to the second-order optical
susceptibility $\chi_{abc}^{(2)}(-\omega,\omega,0)$\cite{Hughes1996}.
In the zero frequency limit,
\begin{equation}
r_{abc}(0)=-\frac{2}{\varepsilon_a(0)\varepsilon_b(0)}\lim_{\omega\rightarrow 0}\chi_{abc}^{(2)}(-2\omega,\omega,\omega).
\end{equation}
Furthermore, for the photon energy $\hbar\omega$ well below the band gap,
$\chi_{abc}^{(2)}(-2\omega,\omega,\omega)$ and $n(\omega)$ are
nearly constant. In this case, the linear electro-optic coefficient
$r_{abc}(\omega)\approx r_{abc}(0)$\cite{wang2015,guo2004}.

To obtain numerically accurate optical properties, we perform test calculations for selenium and tellurium with several different $k$-point meshes
until the calculated optical properties converge to a few percent. As a result, dense $k$-point meshes of 40$\times$40$\times$36
and 50$\times$50$\times$38 are adopted here for selenium and tellurium, respectively. Furthermore, to ensure that $\varepsilon'$
and $\chi'^{(2)}$ obtained by the Kramer-Kronig transformation are reliable, about 27 bands per atom are adopted in the optical calculations.
The function $\delta$ in Eqs. (1), (4) and (5) are approximated by a Gaussian function with $\Gamma=0.2$ eV.

It is well known that the band gap of a semiconductor is generally under-estimated by the local density approximation (LDA)
and GGA calculations (see, e.g., Table II below) where many-body effects especially quasiparticle self-energy correction
are not adequately taken into account.
On the other hand, Eqs. (1), (4) and (5) indicate that correct band gaps would be important
for obtaining accurate optical properties. Therefore, we further perform the band structure calculations using the hybrid
Heyd-Scuseria-Ernzerhof (HSE) functional~\cite{heyd2003} which is known to produce much improved band gaps for semiconductors.
We then take the self-energy corrections into account by the so-called scissors correction (SC)~\cite{levine_prb_1991}
using either the accurate band gaps from the HSE calculations or simply the available experimental band gaps.
In the SC calculation, the conduction bands are unformly upshifted so that the band gap would match
either the experimental gap or the HSE gap together with the renormalized velocity matrix elements~\cite{levine_prb_1991}.
Indeed, such SC calculations were shown to give rise to the second order nonlinear susceptibility at zero frequency
for low-dimensional materials such as trigonal selenium~\cite{zhong1993}
and graphene-like BN sheet that agree well with the experimental one~\cite{wang2015}.

\section{RESULTS AND DISCUSSION}
\subsection{Electronic band structure}

\begin{table}[htbp]
\begin{center}
\caption{
Calculated ($E_{g}^{GGA}$ and $E_{g}^{HSE-SOC}$) and experimental band gap ($E_{g}^{Exp}$)
as well as scissors operator ($\Delta E_g = E_{g}^{HSE-SOC} - E_{g}^{GGA}$) for  selenium and tellurium.
The values in brackets are from the HSE calculations without the SOC.
}
\begin{ruledtabular}
\begin{tabular}{ccccc}
& $E_{g}^{GGA}$ (eV) &  $E_{g}^{HSE-SOC}$ (eV) &  $E_{g}^{Exp}$ (eV)& $\Delta E_g$ (eV)   \\
\hline
  Se  & 1.002 & 1.735 (1.759) & 2.0\footnotemark[1] & 0.733  \\
  Te  & 0.113 & 0.322 (0.546) & 0.323\footnotemark[2] & 0.209 \\
\end{tabular}
\end{ruledtabular}
\footnotemark[1]{Experimental value from reference~\onlinecite{tutihasi1967}.}

\footnotemark[2]{Experimental value from reference~\onlinecite{anzin1977}.}
\end{center}
\end{table}

Because selenium and tellurium have the same crystalline structures, the band structure of selenium is very similar to that of the tellurium,
as shown in Fig. 2. The two materials are indirect bandgap semiconductors which have the conduction band minimum (CBM) at the H point.
For selenium, the valence band maximum (VBM) is located at the L point, whereas the VBM of tellurium is close to the H point
along the H-K direction. The calculated band gap is 1.002 (0.113) eV in Se (Te), which is significantly smaller than the experimental values
of 2.0 (0.323) eV\cite{tutihasi1967,anzin1977}.
The ratio between the interchain and intrachain distances of tellurium, as mentioned previously, is smaller than that of selenium.
Therefore, the enhanced interchain interaction in tellurium brings about more electrons transferred
from lone pair states to antibonding states and this weakens the Peierls distortion. This explains why the band gap of tellurium
is smaller than that of selenium, even though they have the same crystalline structure.
By using the molecular orbital picture, we can further interpret the energy band as follows.
There are three groups of valence bands which correspond to the $s$ bonding, $p$ bonding and $p$ lone pair states,
and the three lowest conduction bands arise from the $p$ antibonding states. The calculated
band structures of Se and Te agree quite well with previous calculations\cite{Hirayama2015}.
The minor differences stem from the fact that the present GGA calculations are done without SOC.

\begin{figure}[htb]
\begin{center}
\includegraphics[width=8.5cm]{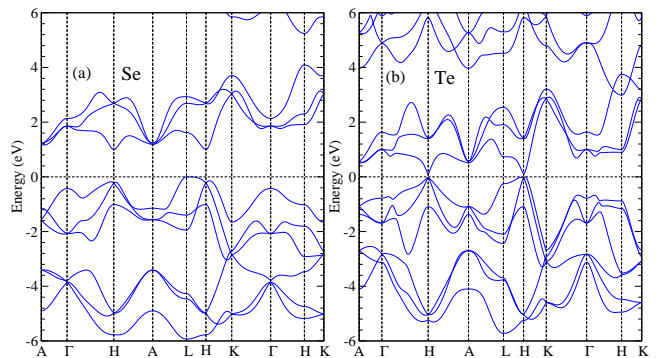}
\end{center}
\caption{Band structures of (a) selenium and (b) tellurium from the GGA calculations.  Both materials possess an indirect band gap.
The top of the valence bands is at 0 eV.}
\end{figure}


It is clear from Eqs. (1), (4) and (5) that the smaller the gap, the larger the magnitude of the dielectric function
and second-order NLO susceptibility. Since the GGA calculation significantly underestimates the energy gap by nearly 50 $\%$,
the dielectric function and second-order NLO susceptibility would be overestimated by the GGA calculations.
Therefore, we also calculate the band structures by using the HSE functional\cite{heyd2003}
in order to get more accurate band gaps. The theoretical band gaps from the  GGA and HSE calculations
together with the experimental values are listed in Table II. Indeed, the band gaps of selenium and tellurium
from the HSE calculations with the SOC included, are in good agreement with the corresponding experimental values (Table II).
Therefore we simply use the energy differences between the HSE calculations with the SOC and GGA band gaps
as the scissors correction energy within the scissors correction scheme\cite{levine_prb_1991},
to obtain accurate linear and nonlinear optical properties for both selenium and tellurium.

\begin{figure}[htb]
\begin{center}
\includegraphics[width=7.0cm]{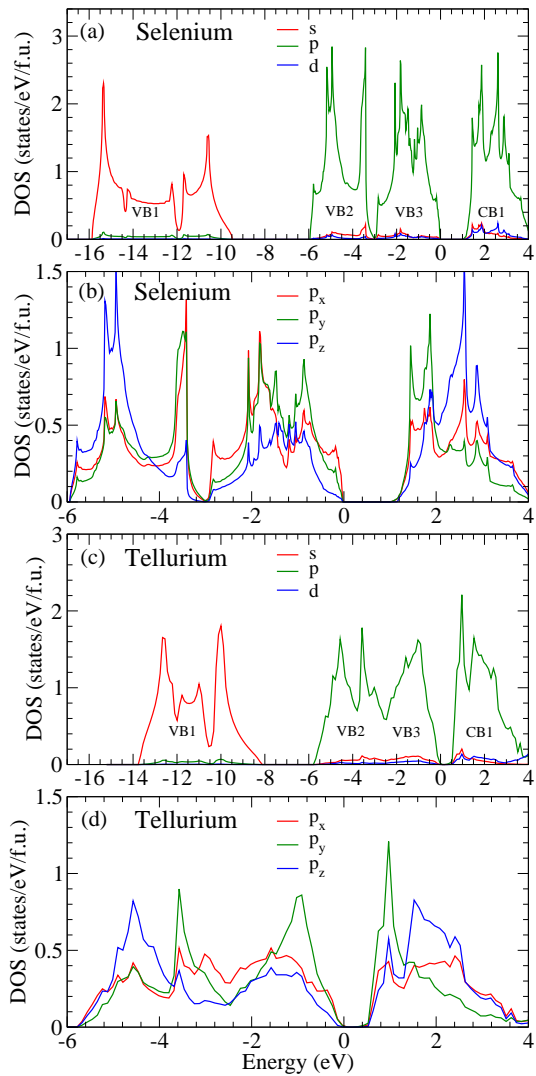}
\end{center}
\caption{Total and orbital-projected densities of states (DOS) for selenium (a) and (b) as well as tellurium (c) and (d).}
\end{figure}

We also calculate total and orbital-projected densities of states (DOS) for selenium and tellurium. Overall,
the DOS spectra of  the two systems are rather similar. Hence, in what follows, we only analyze the DOS spectra of selenium
as displayed in Fig. 3(a). We can see from Fig. 3(a) that the lowest valence bands of selenium ranging
from -15.9 to -9.5 eV stem mainly from the $s$ orbitals.
The top valence bands ranging from -5.9 to 0.0 eV and the lower conduction bands ranging from 1.0 to 4.0 eV, on the other hand,
originate primarily from the $p$ orbitals.

Furthermore, we can analyze the contributions of the orbital-projected DOSs from Fig. 3(b). For example, the lower valence bands
ranging from -5.9  to -4.2 eV and also the conduction bands ranging from 1.9 to 3.3 eV are dominated by the $p_z$ orbital
with some contributions from the $p_x$ and $p_y$ orbitals. The top valence bands ranging from -4.2 to 0.0 eV are of
mainly $p_x$ and $p_y$  orbitals with a certain $p_z$ component. Furthermore, the lower conduction bands ranging
from 1.2 to 1.9 eV mainly consist of $p_y$ orbitals.

\subsection{Linear optical property}

The calculated imaginary (absorptive) part of the optical dielectric function $\varepsilon(\omega)$ 
for selenium and tellurium are shown in Fig. 4. Trigonal selenium and tellurium are of uniaxial 
crystalline structures due to their $D_3$ point group symmetry.
Therefore, their optical properties depend significantly on the light polarization direction. 
Consequently, the imaginary part of the dielectric function for both systems consist of 
two distinctly different components, i.e., light polarization parallel ($E \parallel c$)
and perpendicular ($E \parallel a$) to the $c$ axis (the principal optical axis). Furthermore,
the two systems exhibit strong optical anisotropy which arises from the unique feature of their structures,
i.e., the spiral chains are oriented along the $c$-axis with relatively strong covalent bonds
along each helix but with weak interchain van der waals forces. As Fig. 4(a) shows, the absorptive part
of $E \parallel a$ is much smaller than that of $E \parallel c$ in the low energy range
(about 2.3$\sim$8.8 eV), while it is the opposite in the energy range above 8.8 eV.
Although the similar phenomena are observed in tellurium, it possesses weaker optical anisotropy
than selenium (Fig. 4). This could be explained by the fact that trigonal tellurium has weaker structural anisotropy.

\begin{figure}[htb]
\begin{center}
\includegraphics[width=8.5cm]{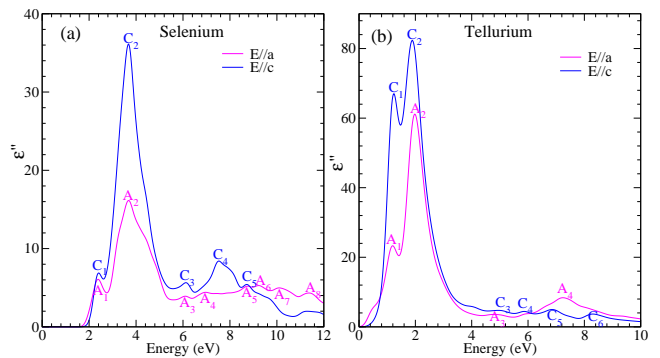}
\end{center}
\caption{The absorptive part of optical dielectric function $\varepsilon"(\omega)$ of (a) selenium and (b) tellurium
for both light polarization perpendicular (E$\parallel$a) and parallel (E$\parallel$c) to the $c$ axis.
}
\end{figure}

Figure 4(a) shows that there are two pronounced peaks in the imaginary part of the dielectric function for $E \parallel c$,
namely, a large peak ($C_2$) located at $\sim$3.7 eV and a relatively small one ($C_4$) in the neighborhood of 7.5 eV.
Furthermore, there are three shoulder peaks at $\sim$2.4 eV ($C_1$) , $\sim$6.1 eV ($C_3$) and $\sim$8.7 eV($C_5$), respectively.
For $E \parallel a$, our theoretical calculations for selenium exhibit a prominent peak ($A_2$) near 3.7 eV.
Beyond that, for $E \parallel a$, the spectrum shows multiple faint bumps ($A_3\sim A_8$). The amplitude of the bumps
above $\sim$5.6 eV increases and then decreases with the photon energy. At 5.6 eV occurs
a deep minimum deriving from the fact that transition from the upper valence triplet to
the lower conduction triplet are already exhausted. The amplitude of the bumps increases 
within the energy range between 5.6 eV and 9.2 eV
due to the transitions from the lower valence triplet to the lower conduction triplet
and also from the upper valence triplet to the upper conduction triplet. For Te, Fig. 4(b) shows
that the spectra of the imaginary part of the dielectric function could be divided into two regions.
In the low-energy region (about 0-4 eV), there are two prominent peaks for
both $E \parallel a$ ($A_1$, $A_2$) and $E \parallel c$ ($C_1$, $C_2$).
The larger peak ($A_2$) occurs at $\sim$2.0 eV and the smaller peak ($A_1$) is found at 1.2 eV for $E \parallel a$.
For $E \parallel c$, the larger ($C_2$) and smaller ($C_1$) peaks are located at 1.9 eV and $\sim$1.2 eV, respectively.
In the high-energy region (about 4-9 eV), a broad peak ($A_4$) centered at $\sim$7.2 eV for $E \parallel a$ exists.
However, the spectrum of $E \parallel c$ exhibits some steadily oscillatory bulges in the energy region.
Therefore, the transitions from the lower valence triplet to the lower conduction triplet and from the upper valence
triplet to the upper conduction triplet are more pronounced.

\begin{figure}[htb]
\begin{center}
\includegraphics[width=8.5cm]{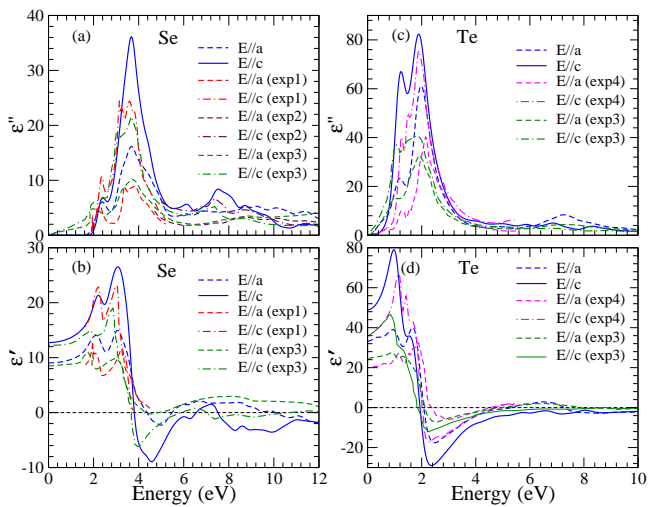}
\end{center}
\caption{Calculated and experimental imaginary ($\varepsilon"(\omega)$) and real part $\varepsilon'(\omega)$
of the dielectric function for (a) and (b) selenium as well as (c) and (d) tellurium for both
light polarization perpendicular (E$\parallel$a) and parallel (E$\parallel$c) to the $c$ axis.
Red, maroon, green and magenta dashed and dot-dashed lines
denote the measured dielectric function spectra from Refs. 31 (exp1), 53 (exp2), 54 (exp3) and 55 (exp4), respectively.}
\end{figure}

\begin{figure}[htb]
\begin{center}
\includegraphics[width=7.0cm]{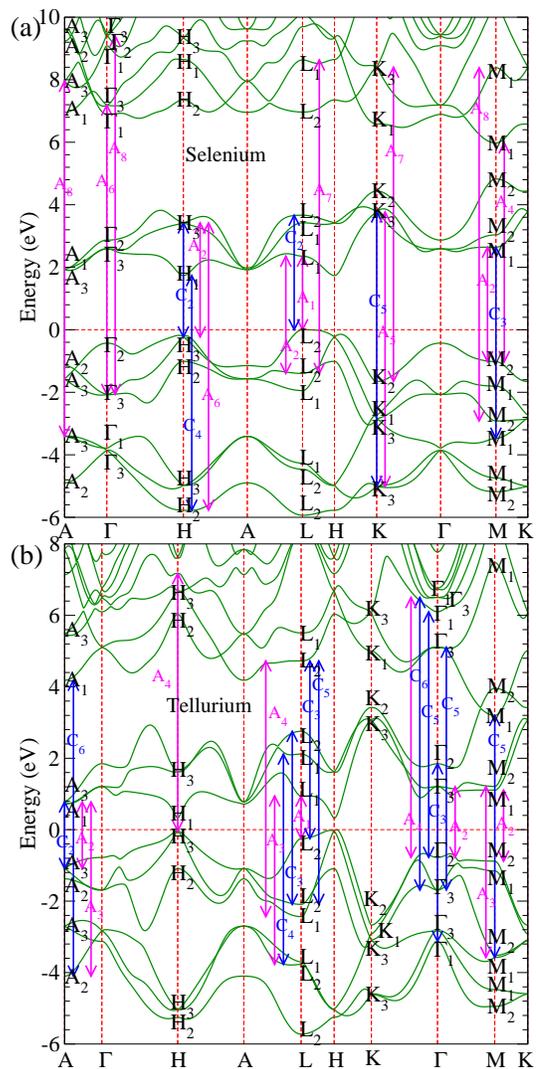}
\end{center}
\caption{(a) Selenium and (b) tellurium band structures from the GGA calculation with the scissors correction.
The symmetries of band states at six high symmetry points are labelled according to the irreducible representations
of the point groups (Table III). The principal interband transitions and the peaks
in the imaginary part of the dielectric function in Fig. 4 are indicated by blue
[light polarized parallel (E$\parallel$c) to the $c$ axis] and pink [light polarized perpendicular
(E$\parallel$a) to the $c$ axis] arrows. The top of the valence bands is at 0 eV.}
\end{figure}

Our calculated static dielectric constants together with the experimental values~\cite{Palik1985} are listed in Table VII.
It is clear from Table VII that the dielectric constants of both selenium and tellurium from the GGA calculations
with the scissors corrections agree very well with the measured ones while, as expected, that obtained from
the GGA calculations without the scissors correction are too large.
Earlier papers\cite{sandrock1971,bammes1972} have reported
the dielectric functions for trigonal selenium and tellurium measured as a function of
photon energy. Therefore, our calculations
are compared with the measurements, as shown in Fig. 5. The imaginary part of the dielectric function would allow us
to find out the information on the major electronic transitions. In what follows, we will compare
the spectra of the imaginary part of the calculated and experimental dielectric functions.
Firstly, for selenium, we find that the positions of the $A_2$ and $C_2$ peaks
at $\sim$3.7 eV are consistent with the experiments\cite{bammes1972}.
Furthermore, small bumps $C_3$ and $C_5$ are also observed in the corresponding experimental spectra\cite{sandrock1971}.
In particular, the position of the $C_5$  bump is consistent with the experiment\cite{sandrock1971}.
Moreover, Fig. 5(a) shows that our calculated five small bumps ($A_3$, $A_4$, $A_5$, $A_7$, $A_8$)
agree reasonably well with that of the measurements\cite{sandrock1971},
other than having the slightly larger amplitude. The experiment and our theoretical calculations
have common peaks $A_1$, $C_1$ and $C_4$, although the energy position of the theoretical peaks is slightly blue-shifted
compared to that of the experimental ones\cite{sandrock1971,bammes1972}. Finally, the positions
of peaks $A_1$ and $C_1$ are consistent with the experiments\cite{tutihasi1967}.

For tellurium, the positions of peaks $A_1$, $A_2$, $A_3$, $A_4$, $C_1$ and $C_2$ in the imaginary part of the dielectric function
agree well with that of the experimental spectra\cite{bammes1972,Tutihashi1969}.
Furthermore, the small $C_3$, $C_5$ and $C_6$ peaks can also be observed in the experimental\cite{bammes1972}
spectra, other than being red-shifted by about 0.2 eV, 0.2 eV and 0.4 eV, respectively.
In the high energy region (4.0-9.0 eV), both the calculated and experimental\cite{bammes1972}
spectra exhibit oscillatory peaks, although the positions of the peaks differ slightly. Overall, the present theoretical
imaginary part of the dielectric function for both materials are in good  agreement with the experiments.
Furthermore, Figs. 5(b) and 5(d) show that the spectral shape of the real part of the dielectric function
agree rather well with that of the experiments\cite{bammes1972,Tutihashi1969}.

\begin{table}[htbp]
\begin{center}
\caption{
Symmetry adapted basis functions of the point groups for six high symmetry $k$-points
in the hexagonal brillouin zone [see Fig. 1(c)].}
\begin{ruledtabular}
\begin{tabular}{ccccccc}
  group & $k$-point & $A_{1}$ & $A_{2}$ & $E$ & $A$ & $B$ \\  \hline
  $D_{3}$ & $A,\Gamma,H,K$ & $s,d_{z^{2}}$ & $p_{z}$ & $p_{x},p_{y},$ &    &     \\
       &    &    &     & $d_{xz},d_{yz}, $  &   &   \\
       &    &    &     & $d_{xy},d_{x^{2}-y^{2}}$   &   &   \\
  $C_{2}$ & $L,M$ &      &     &     & $p_{z},d_{xy},$ & $p_{x},p_{y},$ \\
         &        &   &   &   & $d_{z^{2}},d_{x^{2}-y^{2}}$ & $d_{xz},d_{yz}$ \\
\end{tabular}
\end{ruledtabular}
\end{center}
\end{table}

\begin{table}[htbp]
\begin{center}
\caption{
Dipole selection rules between the band states at six high symmetry $k$-points
in the hexagonal brillouin zone [see Fig. 1(c)].}
\begin{ruledtabular}
\begin{tabular}{cccccc}
      & $E \perp c$ & $E \parallel c$ &     & $E \perp c$ & $E \parallel c$ \\ \hline
  $D_{3}$ & $\Gamma_{1} \longleftrightarrow \Gamma_{3}$ & $\Gamma_{1} \longleftrightarrow \Gamma_{2}$ & $C_{2}$ & $\Gamma_{1} \longleftrightarrow \Gamma_{2}$ & $\Gamma_{1} \longleftrightarrow \Gamma_{1}$ \\
  $\Gamma $ & $\Gamma_{2} \longleftrightarrow \Gamma_{3}$ & $\Gamma_{3} \longleftrightarrow \Gamma_{3}$ & $L$ & $\Gamma_{1} \longleftrightarrow \Gamma_{3}$ & $\Gamma_{2} \longleftrightarrow \Gamma_{2}$ \\
  $(A,H,K)$ & $\Gamma_{3} \longleftrightarrow \Gamma_{3}$ &  & $(M)$ & $\Gamma_{2} \longleftrightarrow \Gamma_{3}$ & $\Gamma_{3} \longleftrightarrow \Gamma_{3}$ \\
\end{tabular}
\end{ruledtabular}
\end{center}
\end{table}

\begin{table}[htbp]
\begin{center}
\caption{
The principal peaks in the imaginary part spectra of the dielectric function
[see Fig. 4(a)] and corresponding direct interband transitions at six high symmetry $k$-points [see Fig. 6(a)] for selenium.}
\begin{ruledtabular}
\begin{tabular}{cccc}
 peak   & direct transitions & peak & direct transitions \\ \hline
$C_{1}$ &                              & $A_{1}$ & $L:9\longrightarrow10$  \\
$C_{2}$ & $H:8,9\longrightarrow11,12;$ & $A_{2}$ & $H:8,9\longrightarrow11,12;$ \\
        & $L:9\longrightarrow12$       &         & $L:8\longrightarrow10;$ \\
        &                              &         & $M:9\longrightarrow10$  \\
$C_{3}$ & $M:6\longrightarrow10$       & $A_{3}$ &                         \\
$C_{4}$ & $H:4\longrightarrow10$       & $A_{4}$ & $M:9\longrightarrow13$ \\
$C_{5}$ & $K:4,5\longrightarrow10,11$  & $A_{5}$ & $K:4,5\longrightarrow10,11$ \\
        &                              & $A_{6}$ &  $\Gamma:7,8\longrightarrow14,15;$ \\
        &                              &         &  $H:4\longrightarrow11,12$ \\
        &                              & $A_{7}$ &  $K:9\longrightarrow14,15;$ \\
        &                              &         &  $L:8\longrightarrow14$ \\
        &                              & $A_{8}$ &  $A:3\longrightarrow10,11;$ \\
        &                              &         &  $A:5,6\longrightarrow14,15;$ \\
        &                              &         &  $\Gamma:7,8\longrightarrow17;$ \\
        &                              &         &  $M:7\longrightarrow14$ \\
\end{tabular}
\end{ruledtabular}
\end{center}
\end{table}

The electronic band structure and linear optical properties of selenium and tellurium have been
studied both theoretically\cite{mcfee1970,sandrock1971,isomaki1980,Treusch1980}
and experimentally\cite{tutihasi1967,roberts1968,anzin1977,sandrock1971,bammes1972} before.
Nevertheless, no detailed analysis on the main peaks in the imaginary part of the dielectric function
in term of interband transitions has been reported. Here we perform such a detailed analysis.
Equations (1), (4) and (5) show that the imaginary parts of the dielectric function and second-order NLO susceptibility
are closely connected with the dipole-allowed interband transitions. Therefore, we can understand the origin of
the peaks in the imaginary part of the dielectric function by analyzing the symmetry of the band states
and also considering the dipole transition selection rules. First, by using the projection method of the group theory,
we deduce the symmetry-adapted basis functions in terms of the atomic orbitals (Table III).
Second, we determine the symmetry of the band states (Fig. 6) for six principal symmetry points ($A, \Gamma, H, L, K, M$)
by means of the comparison between the symmetry-adapted basis functions and the calculated orbital characters
of the band states at the six symmetry points. It should be noted that the atomic configurations
of selenium and tellurium are $4s^2$, $4p^4$ and $5s^2$, $5p^4$. Therefore,
we only consider the symmetry-adapted basis functions of $s$, $p_x$, $p_y$ and $p_z$ orbitals.
Such deduced symmetries at the $A, \Gamma, H,$ and $K$ points are consistent with the results of previous calculation\cite{isomaki1980}.
After considering the selection rules (Table IV)\cite{Treusch1980},
we could assign the peaks in the imaginary part of the dielectric function to the direct interband transitions
at the six symmetry points, as shown in Fig. 6 as well as in Tables V and VI. For example, for trigonal selenium,
the $A_2$ peak at $\sim$3.7 eV [see Fig. 4(a)] stems from transitions from the $H_3$ state at the top of valence band to
the conduction band state $H_3$ ($\sim$ 3.4 eV) at $H$-point and from the $L_2$ valence band state at -1.4 eV
to the $L_1$ state at the bottom of conduction band at the $L$-point. Furthermore, the $A_2$ peak is
associated with the transition from the $M_2$ state at the top of valence band to the  $M_1$ state
at the bottom of conduction band at $M$-point. The $C_2$ peak at $\sim$3.7 eV could be attributed to transitions starting
from the $H_3$ state at the top of valence band to the conduction band states $H_3$ ($\sim$3.4 eV) at $H$-point
and from the $L_2$ state at the top of valence band to the $L_2$ conduction band state ($\sim$3.7 eV) at the $L$-point.

\begin{table}[htbp]
\begin{center}
\caption{
The principal peaks in the imaginary part spectra of the dielectric function [see Fig. 4(b)] and corresponding
direct interband transitions at six high symmetry $k$-points [see Fig. 6(b)] for tellurium.}
\begin{ruledtabular}
\begin{tabular}{cccc}
peak & direct transitions & peak & direct transitions \\ \hline
$C_{1}$ &                                    & $A_{1}$  & $L:9\longrightarrow10$  \\
$C_{2}$ & $A:8,9\longrightarrow11,12$        & $A_{2}$  & $A:8,9\longrightarrow11,12;$ \\
        &                                    &          & $\Gamma:9\longrightarrow10,11;$ \\
        &                                    &          & $M:9\longrightarrow10$  \\
$C_{3}$ & $\Gamma:4\longrightarrow12;$       & $A_{3}$  & $A:4\longrightarrow11,12;$ \\
        & $L:8\longrightarrow12;$            &          &  $L:5\longrightarrow10;$ \\
        & $L:9\longrightarrow13$             &          &  $M:6\longrightarrow11$  \\
$C_{4}$ & $L:6\longrightarrow11$             &  $A_{4}$ & $\Gamma:9\longrightarrow16,17;$ \\
        &                                    &          & $H:8,9\longrightarrow16;$ \\                              &                                    &          &  $L:7\longrightarrow13$ \\
$C_{5}$ & $\Gamma:7,8\longrightarrow13,14;$  &          &                         \\
        & $\Gamma:9\longrightarrow15;$       &          &                         \\
        & $M:6\longrightarrow12;$            &          &                         \\
        & $L:8\longrightarrow13$             &          &                         \\
$C_{6}$ & $A:4\longrightarrow13;$            &          &                         \\
        & $\Gamma:7,8\longrightarrow16,17$   &          &                          \\
\end{tabular}
\end{ruledtabular}
\end{center}
\end{table}

Similarly, for tellurium, we can assign the $A_2$ peak [see Fig. 4(b)] to the transition from the $\Gamma_2$ state at the top
of valence band to the $\Gamma_3$  state at the bottom of conduction band at the $\Gamma$-point and from the $M_2$ state
at the top of valence band to the $M_1$ state at the bottom of conduction band at the $M$-point.
It is noted that the transition from the $A_3$ state at the top of the valence band to the conduction
band state $A_3$ ($\sim$0.8 eV) at $A$-point contributes to the $A_2$ and $C_2$ peaks.

\subsection{Second harmonic generation and linear electro-optic coefficient}

Point group $D_3$ has ten nonvanishing second-order NLO susceptibility elements.\cite{Shen1984,Boyd2003} 
However, the crystalline symmetry $D_3^4$ of selenium and tellurium further reduces this number to five. 
Our {\it ab initio} calculations show that there are only two independent
elements among the five nonzero elements because $\chi^{(2)}_{xxx}=-\chi^{(2)}_{xyy}=-\chi^{(2)}_{yxy}$, $\chi^{(2)}_{xyz}=-\chi^{(2)}_{yzx}$. 
Our calculations for selenium also reveal that
the values of the $\chi^{(2)}_{xxx}$ element of the two helical structures ($D_3^4$ and $D_3^6$) are equal but
the values of the $\chi^{(2)}_{xyz}$ element differ in sign. 
Therefore, here we present only $\chi^{(2)}_{xxx}$ and $\chi^{(2)}_{xyz}$ for space group $D_3^4$,
as mentioned above in Sec. II. 
The calculated real and imaginary parts as well as the absolutes values
of these two elements are displayed in Fig. 7 and Fig. 8 for selenium and tellurium, respectigvely.
We note that with the scissor corrections, the line shapes of the calculated NLO spectra are hardly changed
and thus only the NLO spectra calculated with the scissor corrections are displayed in Figs. 7 and 8.
However, the peak positions are blue-shifted by about the energy of the scissors correction ($\Delta E_g$).
Moreover, the magnitude of the second-order susceptibility gets reduced (see Table VII).

\begin{table*}[htbp]
\begin{center}
\caption{
Static dielectric constants ($\varepsilon_x = \varepsilon_y$ and $\varepsilon_z$),
second-order susceptibility $\chi^{(2)}_{xxx}(0)$ (pm/V), $\chi^{(2)}_{xyz}(0)$ (pm/V)
and $|\chi^{(2)}_{xxx}(0.113 $eV$)|$ (pm/V) as well as  LEO coefficient $r_{xxx}(0)$ (pm/V)
and $r_{xyz}(0)$ (pm/V) of selenium and tellurium calculated without (GGA) and with (SC) scissors correction.
Available experimental data and previous calculation are also listed for comparison.}
\begin{ruledtabular}
\begin{tabular}{ccccccccc}
        &  & $\varepsilon_x$ & $\varepsilon_z$ & $\chi^{(2)}_{xxx}(0)$  & $\chi^{(2)}_{xyz}(0)$ & $r_{xxx}(0)$ & $r_{xyz}(0)$ & $|\chi^{(2)}_{xxx}(0.113 $eV$)|$  \\ \hline
 Se& GGA & 10.8 & 15.4 & 330, 440\textsuperscript{\emph{a}} & 6   & -5.63   & -0.10   &  335   \\
   & SC & 9.0  & 12.7 & 145, 194\textsuperscript{\emph{a}} &  5  & -3.58   & -0.12   &  146  \\
  & exp.& 6.2$\sim$8.4\textsuperscript{\emph{b}}  & 12.7\textsuperscript{\emph{b}} &    &    &    &    & 159$\pm$84\textsuperscript{\emph{c}} \\
        &   &                                           &    &    &    &    &    & 194$\pm$50\textsuperscript{\emph{d}} \\
Te& GGA & 40.6 & 57.2 &  -2190  & 7927   & 2.66   & -9.63   & 3163    \\
        & SC & 33.2 & 49.0 & 169   & 1009   &  -0.30 & -1.82   &  475   \\
        & exp.& 33.0\textsuperscript{\emph{b}} & 54.0\textsuperscript{\emph{b}} &    &    &    &    & 1840$\pm$560\textsuperscript{\emph{d} } \\
  &    &    &    &    &    &    &    & 1843$\pm$586\textsuperscript{\emph{e} } \\
\end{tabular}
\end{ruledtabular}
\footnotemark[1]{LDA calculations with the SC from reference~\onlinecite{zhong1993}.}

\footnotemark[2]{Experimental value from reference~\onlinecite{Palik1985}.}

\footnotemark[3]{Experimental value from reference~\onlinecite{patel1966}.}

\footnotemark[4]{Experimental value from reference~\onlinecite{day1971}.}

\footnotemark[5]{Experimental value from reference~\onlinecite{mcfee1970}.}
\end{center}
\end{table*}

\begin{figure}[htb]
\begin{center}
\includegraphics[width=8cm]{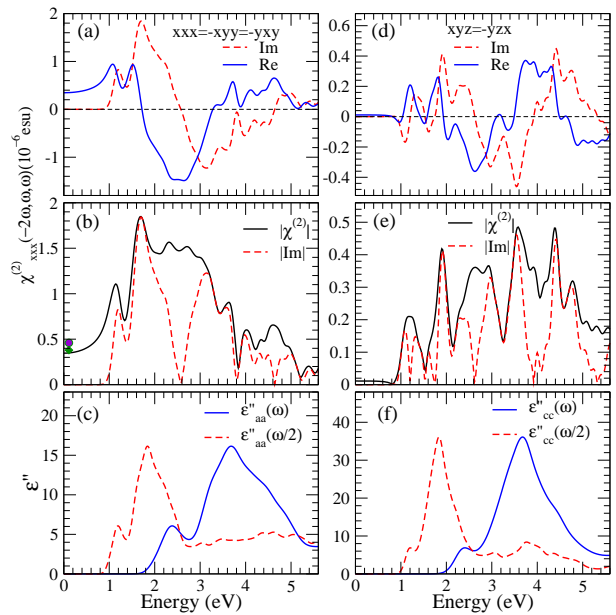}
\end{center}
\caption{(a) and (d) Real and imaginary parts as well as (b) and (e) absolute value of the second-order susceptibility
(a) and (b) $\chi^{(2)}_{xxx}$ as well as (d) and (e) $\chi^{(2)}_{xyz}$ of selenium.
(c) and (f) Imaginary part $\varepsilon"(\omega)$ of the dielectric function for
light polarization perpendicular and parallel to the $c$ axis, respectively.
In (b), the green diamond and violet circle denote the experimental SHG values from
Refs. 30 and 33, respectively. }
\end{figure}

In Table VII, we list the calculated static dielectric constant $\varepsilon(0)$, second-order NLO susceptibility $\chi^{(2)}(0,0,0)$
and zero frequency LEO coefficient $r(0)$. Note that the $r(0)$ and $\chi^{(2)}(0,0,0)$ are listed in the units
of  SI pm/V unit and 1 pm/V ($=\frac{3}{4\pi}\times10^{-8}$ esu).
Interestingly, Table VII shows that tellurium exhibits large static second-order NLO susceptibility, especially $\chi^{(2)}_{xyz}(0)$,
which is up to 100 times larger than that of GaN in both zinc-blende and wurtzite structure\cite{Gavrilenko2000,Cai2009}.
Furthermore, for both materials, the static second-order NLO susceptibility exhibits strong anisotropy. 
The same phenomenon is observed in the LEO coefficient.
Moreover, selenium has large LEO coefficient $r_{xxx}(0)$ ($\sim$3.6 pm/V),
being more than six times larger than that of bulk GaN polytypes\cite{Gavrilenko2000,Cai2009}.

Let us now compare our calculated $|\chi^{(2)}_{xxx}(0)|$ and $|\chi^{(2)}_{xxx}(0.113)|$ with
the previous LDA calculations and available experiments. Firstly, our calculated
$|\chi^{(2)}_{xxx}(0.113)|$ for selenium agrees rather well with the experimental values reported in ref.\cite{patel1966}
and also ref. \cite{day1971} [see Table VII and Fig. 7(b)].
Secondly, for $|\chi^{(2)}_{xxx}(0)|$, the agreement between our
GGA calculations and the previous LDA calculations \cite{zhong1993}
is also rather good (see Table VII). For tellurium, Table VII and Fig. 8(b) show that
our calculated $|\chi^{(2)}_{xxx}(0.113)|$ is much smaller than the available experiment values\cite{mcfee1970,day1971}.
Nonetheless, we note that in the energy range of 0.0$\sim$0.3 eV, $|\chi^{(2)}_{xxx}|$ increases
rapidly with the photon energy and the energy of 0.113 eV sits right at the middle of the
steep slope [Fig. 8(b)]. Thus, a small error in the energy position could cause
a large discrepancy between the theory and experiment. Indeed a small red-shift ($\sim$0.1 eV) could
bring the calculated and experimental values of $|\chi^{(2)}_{xxx}(0.113)|$ in good agreement.

\begin{figure}[htb]
\begin{center}
\includegraphics[width=8cm]{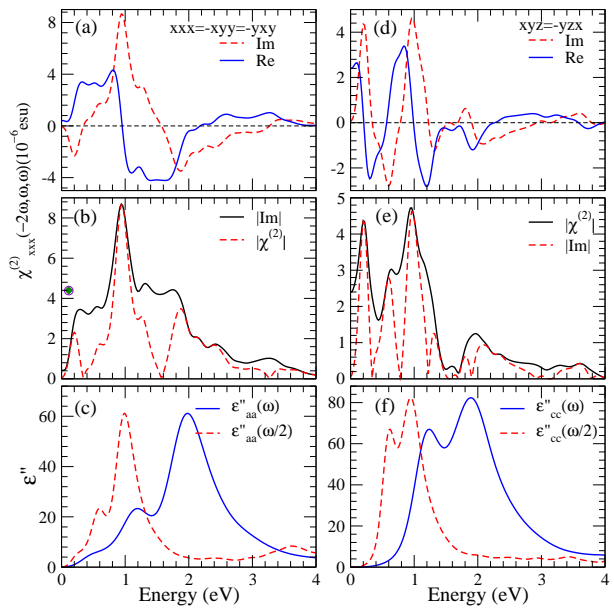}
\end{center}
\caption{(a) and (d) Real and imaginary parts as well as (b) and (e) absolute value of the second-order susceptibility
(a) and (b) $\chi^{(2)}_{xxx}$ as well as (d) and (e) $\chi^{(2)}_{xyz}$ of tellurium. 
(c) and (f) Imaginary part $\varepsilon"(\omega)$ of the dielectric function for
light polarization perpendicular and parallel to the $c$ axis, respectively.
In (b), the green diamond and violet circle denote the experimental SHG values
from Refs. 33 and 34, respectively.}
\end{figure}

The SHG involves not only single-photon ($\omega$) resonance but also double-photon (2$\omega$) resonance.
Therefore, to further analyze the NLO responses, we plot the modulus of the imaginary parts of the second-order
susceptibility $|\chi^{(2)}(-2\omega,\omega,\omega)|$ as well as dielectric
functions $\varepsilon''(\omega)$ and $\varepsilon''(\omega/2)$ together in Figs. 7 and 8
for selenium and tellurium, respectively, in order to understand the prominent features in the spectra
of $\chi^{(2)}(-2\omega,\omega,\omega)$. For selenium, Fig. 7 shows that
the threshold of the $\chi^{(2)}(-2\omega,\omega,\omega)$ spectra and also the absorption edge
of $\varepsilon''(\omega/2)$ is at $\sim$0.87 eV ($\frac{1}{2}E_{g}$),
while the absorption edge of $\varepsilon''(\omega)$ is at $\sim$1.73 eV ($E_{g}$).
Therefore, the SHG spectra can be divided into two parts. The first part from 0.87 to 2.50 eV
stems predominantly from double-photon resonances. The second part (above 2.5 eV) is mainly associated
single-photon resonances with some contribution from double-photon resonances [see Figs. 7(c) and 7(f)].
These two types of resonances cause the SHG spectra to oscillate
and decrease gradually in the higher energy region. Figure 7 indicates that for selenium, both the real and imaginary parts
of the second-order NLO susceptibility $\chi^{(2)}_{xyz}(-2\omega,\omega,\omega)$ show an oscillatory behavior with the energy.
Indeed, the spectrum of $|\chi^{(2)}_{xyz}(-2\omega,\omega,\omega)|$ oscillates rapidly and has the maximum
of about 0.49$\times$10$^{-6}$ esu at $\sim$3.6 eV [see Fig. 7(e)].
It is clear from Fig. 7(b) that the spectrum of the absolute value
of $\chi^{(2)}_{xxx}(-2\omega,\omega,\omega)$ shows a broad plateau from 1.34 eV to 3.44 eV and reaches the
maximum of 1.85$\times$10$^{-6}$ esu at 1.7 eV which is a few times larger than that of GaN~\cite{Gavrilenko2000,Cai2009},
a widely used NLO semiconductor.

For tellurium, likewise, the spectral structure in the energy range from $\sim$0.16 ($\frac{1}{2}E_{g}$)
to $\sim$1.0 eV ($E_{g}$) is formed mainly by double-photon ($2\omega$) resonances. The rest structure (above $\sim$1.0 eV)
arises mainly from single-photon ($\omega$) resonances with some contribution from $2\omega$ resonances.
It is noted that all the NLO susceptibilities of tellurium are rather large in the photon energy range of 0.0$\sim$3.0 eV.
For example, the magnitude of $|\chi^{(2)}_{xyz}(-2\omega,\omega,\omega)|$ is around 4.72$\times$10$^{-6}$ esu at 0.96 eV.
Interestingly, for both systems, the real part, imaginary part and absolute value of $\chi^{(2)}_{xyz}$
are smaller than $\chi^{(2)}_{xxx}$, and produce relatively pronounced oscillations compared
to $\chi^{(2)}_{xxx}$. The phenomenon is explained by the fact that the two materials are helical chains
along the $c$-axis and possess a high degree of anisotropy.
In particular, the absolute value of second-order NLO susceptibility $\chi^{(2)}_{xxx}$ of tellurium
is as high as 8.69$\times$10$^{-6}$ esu at 0.94 eV, which is nearly sixteen times larger than that of GaN~\cite{Gavrilenko2000,Cai2009}.
This suggests that tellurium would be a superior NLO material and has potential application in NLO and LEO optical devices
such as frequency conversion, optical switching, second-harmonic generation, optical modulation and sensing devices.

\section{CONCLUSION}
To conclude, we have calculated the linear and nonlinear optical properties of trigonal selenium and tellurium
based on the DFT with the GGA. In order to adequately take into account of many-body effects especially
quasiparticle self-energy correction, we further perform the relativistic band structure calculations using the hybrid
Heyd-Scuseria-Ernzerhof functional and use the much improved band gaps to calculate the optical
properties with the scissors corrections. We find that the two materials exhibit large SHG and LEO effects.
Also, their linear and nonlinear optical responses are highly anisotropic due to their structural anisotropy.
In particular, the second-order NLO susceptibilities of tellurium are huge in the photon energy range of 0$\sim$3 eV,
with the magnitudes of $\chi^{(2)}_{xxx}$  being as large as 8.69$\times$10$^{-6}$ esu, which is about sixteen times larger than
that of GaN, a widely used NLO material. Furthermore, tellurium is found to exhibit gigantic static SHG coefficients
with the $\chi^{(2)}_{xyz}$ component being up to 100 times larger than that of GaN.
On the other hand, selenium is shown to possesses large LEO coefficient $r_{xxx}(0)$ which are
more than six times larger than that of GaN polytypes.
Therefore, tellurium and selenium are excellent NLO materials and may find valuable applications
in NLO and LEO devices such as  electro-optical switch, frequency conversion, phase matching and light signal modulator.
Interestingly, our calculations also reveal that for the two different helical structures of each material,
the values of $\chi^{(2)}_{xxx}$ are equal but the values of $\chi^{(2)}_{xyz}$ differ in sign, thus suggesting that
the SHG spectroscopy is a useful probe of the chirality of these helical materials. 
The calculated static dielectric constants and also SHG coefficients at the CO$_2$ laser frequency
are in good agreement with the experiments. Furthermore,
the energy positions and shapes of the principal features in the calculated optical dielectric function spectra
of both materials agree rather well with the available experimental ones. They are also analyzed in terms
of the calculated electronic band structures especially symmetries of the involved band states and
dipole transition selection rules.
The prominent structures in the spectra of  $\chi^{(2)}(-2\omega,\omega,\omega)$ are also related to
single-photon and double-photon resonances.
We believe that our work will stimulate further experiments on the NLO and LEO effects
in these interesting materials.

\section*{ACKNOWLEDGEMENTS}

M. C. thanks Department of Physics and Center for Theoretical Physics, National Taiwan University
for its hospitality during her three months visit there when parts of this work were carried out.
Work at Xiamen University is supported by the National Key R$\&$D Program of China (Grant No. 2016YFA0202601),
and the National Natural Science Foundation of China (No. 11574257).
G. Y. G. acknowledges support from the Ministry of Science and Technology, the Academia Sinica,
the National Center for Theoretical Sciences in Taiwan.

\end{document}